\documentclass[useAMS,usenatbib]{aa}
\usepackage{txfonts,color}
\usepackage[authoryear]{natbib}
\bibpunct{(}{)}{;}{a}{}{,}
\usepackage[dvips]{graphicx}
\newcommand{\simgt}{\lower.5ex\hbox{$\; \buildrel > \over \sim \;$}}
\newcommand{\simlt}{\lower.5ex\hbox{$\; \buildrel < \over \sim \;$}}

\begin{document}

\title{Star-formation efficiency and metal enrichment
  of the intracluster medium in local massive clusters of galaxies}

\author{Yu-Ying Zhang\inst{1,2},
Tatiana F. Lagan\'a\inst{3,1},
Daniele Pierini\inst{4,**},
Ewald Puchwein\inst{5,6},
Peter Schneider\inst{1},
\and
Thomas H. Reiprich\inst{1}
}

\institute{
Argelander-Institut f\"ur Astronomie, Universit\"at Bonn, Auf dem
  H\"ugel 71, 53121 Bonn, Germany
  \and National Astronomical Observatories, Chinese Academy of Sciences,
  Beijing 100012, China
\and Universidade de S\~ao Paulo, Instituto de Astronomia, Geof\'isica e
  Ci\^encias Atmosf\'ericas, Departamento de Astronomia, Rua do Mat\~ao 1226,
  Cidade Universit\'aria, CEP:05508-090, S\~ao Paulo, SP, Brasil
  \and Max-Planck-Institut f\"ur extraterrestrische Physik, Giessenbachstra\ss
  e, 85748 Garching, Germany
\and Heidelberger Institut f\"ur Theoretische Studien, Schloss-Wolfsbrunnenweg
  35, 69118 Heidelberg, Germany
\and Max-Planck-Institut f\"ur Astrophysik, Karl-Schwarzschild-Stra\ss e 1,
  85741 Garching, Germany }

\authorrunning{Zhang et al.}

\titlerunning{Star-formation efficiency and metal enrichment
  of the ICM in local massive clusters of galaxies}

\date{Received 01/03/2011 Accepted 02/09/2011}

\offprints{Y.-Y. Zhang\\
*** Guest astronomer at the MPE}

\abstract{We have investigated the baryon-mass content in a subsample
  of 19 clusters of galaxies extracted from the X-ray flux-limited
  sample HIFLUGCS according to their positions in the sky. For these
  clusters, we measured total masses and characteristic radii on the
  basis of a rich optical spectroscopic data set, the physical
  properties of the intracluster medium (ICM) using \emph{XMM-Newton}
  and \emph{ROSAT} X-ray data, and total (galaxy) stellar masses
  utilizing the SDSS DR7 multi-band imaging. The observed (hot)
  gas-mass fractions are almost constant in this mass range. We
  confirm that the stellar mass fraction decreases as the total mass
  increases and shows ($20\pm4$)\% scatter; in addition, we show that
  it decreases as the central entropy increases. The latter behavior
  supports a twofold interpretation where heating from merging
  quenches the star-formation activity of galaxies in massive systems,
  and feedback from supernovae and/or radio galaxies drives a
  significant amount of gas to the regions beyond $r_{500}$ or,
  alternatively, a substantially large amount of intracluster light
  (ICL) is associated with galaxies in nonrelaxed systems.
  Furthermore, less massive clusters are confirmed to host less gas
  per unit total mass; however, they exhibit higher mass fractions in
  metals, so that their ICM is more metal-rich. This again supports
  the interpretation that in the potential wells of low-mass systems
  the star-formation efficiency of galaxies was high or,
  alternatively, some gas is missing from the hot phase of the
  ICM. The former hypothesis is preferred as the main driver of the
  mass-dependent metal enrichment since the total mass-to-optical
  luminosity ratio increases as the total mass increases.
}

\maketitle

\section{Introduction}

Galaxy clusters are the largest gravitationally bound structures in
the Universe, and they have become an important cosmological probe to
constrain dark energy (e.g., Vikhlinin et al. 2009). Calibrating the
baryon content of galaxy clusters is thus a key input for
understanding the structure-formation history, as well as systematic
uncertainties in constraining cosmology with galaxy clusters as
standard candles.

The hot intracluster medium (ICM; e.g., Sarazin \& Bahcall 1977) and
the stars bound to member galaxies have been considered to be the main
contributors to the baryon content in clusters. The baryon-mass
fraction of a galaxy cluster was thus defined as the sum of the
gas-mass fraction ($f_{\rm gas}:= M_{\rm gas}/M_{\rm tot}$, the
gas-to-total mass ratio) and the stellar mass fraction
($f_{\ast}:=M_{\ast}/M_{\rm tot}$, the stellar-to-total mass ratio in
which the stars in galaxies are considered). We can use X-ray data
alone to measure the gas-mass fraction (e.g., Ettori et al. 2002;
Zhang et al. 2006). There are two well-established methods of
measuring the total stellar mass in a galaxy cluster. One is to
measure the stellar mass for the total galaxy population of a cluster
from the total luminosity based on the galaxy luminosity function
(GLF) and a stellar mass-to-optical light ratio that depends on the
mass of a cluster (e.g., David et al. 1990; Roussel et al. 2000; Lin
et al. 2003). The other is to measure the stellar masses of the
individual member galaxies and to construct the stellar mass function
in order to sum these stellar masses (e.g., see section 3 in Giodini
et al. 2009). The contribution to the total stellar mass budget
associated with intracluster light (ICL; e.g., Zibetti et a. 2005)
could be significant especially on the mass scale of galaxy groups if
the ICL makes more than 10--20\% of the total light (e.g., Gonzalez et
al. 2007). The discrepancies between measurements of the baryon-mass
fraction of galaxy clusters and groups (e.g., Lin et al. 2003;
Gonzalez et al. 2007; Krick \& Bernstein 2007; Lagan\'a et al. 2008,
2011; Giodini et al. 2009; Andreon 2010; Dai et al. 2010) and the
measurement from the Wilkinson Microwave Anisotropy Probe (WMAP)
5-year data (Dunkley et al. 2009) allow for an ICL-to-total stellar
mass ratio in a wide range up to $\sim 50$\% with a controversial mass
dependence. Most observational studies do not support dominant
contributions from the ICL (e.g., Zibetti et al. 2005; see Arnaboldi
\& Gerhard 2010 for a recent review). Furthermore, theoretical studies
reach controversial conclusions on the origin and mass dependence of
the ICL fraction (e.g., Murante et al. 2004; Dolag et al. 2010 and
references therein).

Understanding the baryon-mass fraction and its mass dependence will be
a milestone to understand astrophysics in galaxy clusters, e.g., the
origin of the ICL (e.g., Pierini et al. 2008), star-formation history
(e.g., Fritz et al. 2010), and metal-enrichment history (e.g.,
Kapferer et al. 2009). This also helps to control systematic
uncertainties in high-precision cluster cosmology.

The HIFLUGCS (Reiprich \& B\"ohringer 2002) provides an X-ray
flux-limited sample of 64 nearby ($z<0.1$) clusters selected from the
\emph{ROSAT} All-Sky Survey (RASS). Nineteen clusters in the HIFLUGCS
have been imaged beyond $9r_{500}$\footnote{The cluster radius,
  $r_{\Delta}$, is the radius within which the mass density is
  $\Delta$ times the critical density, $\rho_{\rm
    c}(z)=E^2(z)3H_{0}^2/(8\pi G)$, where
  $E^2(z)=\Omega_{\Lambda}+\Omega_{\rm
    m}(1+z)^3+(1-\Omega_{\Lambda}-\Omega_{\rm m})(1+z)^2$. In this
  work, we measured two cluster radii, $r_{500}$ and $r_{200}$. As
  shown in Lagan\'a et al. (2011), the coverage of $9r_{500}$ is
  required to perform a robust background subtraction to construct the
  GLF of the cluster.} in the Sloan Digital Sky Survey (SDSS, DR7,
Abazajian et al. 2009). Since these 19 clusters were selected from the
HIFLUGCS based on their sky positions, this subsample is unbiased with
respect to the HIFLUGCS. Our sample has therefore inherited the nature
of the flux selection.

In this paper, we analyze \emph{XMM-Newton} and \emph{ROSAT} X-ray
data, as well as optical data for this sample of those 19 clusters, and
investigate the baryon content in clusters and its mass
dependence. This study, similar to Andreon (2010), has the advantage
over many previous studies that the cluster total masses and reference
radii are derived from the velocity dispersion based on optical
spectroscopic data (see Zhang et al. 2011; and references therein)
that are independent of both the X-ray data used to determine the gas
masses and the optical imaging data used to measure the stellar
masses. It enables a relatively objective investigation of the
intrinsic scatter in the gas-mass, stellar mass and baryon-mass
fractions including contributions from the intrinsic scatter of mass
estimates.

We organize the paper as follows. In Sect.~\ref{s:data}, we describe
the sample and data analysis. We present our results in
Sect.~\ref{s:res}, as well as discussions in Sect.~\ref{s:dis}, and
summarize our findings in Sect.~\ref{s:conc}. We assume $\Omega_{\rm
  m}=0.3$, $\Omega_\Lambda=0.7$, and $H_{0} = 70 \rm km~\rm s^{-1}~\rm
Mpc^{-1}$. Throughout this paper we apply the BCES regression fitting
method taking measurement errors into account in both variables and
their covariance (Akritas \& Bershady 1996).  Confidence intervals
correspond to the 68\% confidence level.

\section{Sample and data}
\label{s:data}

\subsection{Sample}
\label{s:sample}

The HIFLUGCS is an X-ray flux-limited sample of 64 nearby ($z<0.1$)
clusters selected from the RASS. Nineteen clusters
(Table~\ref{t:sample}) in the HIFLUGCS have available
\emph{XMM-Newton} and SDSS DR7 data (Abazajian et al. 2009).

\subsection{Cluster total mass and radius}
\label{s:r500}

Recent observations suggest deviations from hydrostatic equilibrium in
galaxy clusters (e.g., Zhang et al. 2008, 2010; Mahdavi et
al. 2008). We therefore avoid using the X-ray hydrostatic mass (e.g.,
Zhang et al. 2009) or caustic mass (e.g., Rines \& Diaferio 2006) as
the cluster mass. Furthermore, to guarantee a fair study, we choose to
obtain the total mass from a quantity that is not linked to either
quantities that we study, i.e., stellar mass and gas mass.

We calculate the cluster mass from the velocity dispersion using the
optical spectroscopic data listed in Table 1 in Zhang et al. (2011;
also see references therein), which are independent of both the X-ray
data used to derive the ICM properties and the optical SDSS imaging
data used to derive the stellar masses. 

Taking the ``harmonic'' velocity dispersion ($\sigma_{\rm a,p}$)
measured within an aperture of 1.2 Abell radii ($a=2.57$~Mpc) from
Zhang et al. (2011), we follow the method described in Sect.~3 in
Biviano et al. (2006) of computing the cluster mass $M_{\sigma}$. We
first determine an initial estimate of the mass from
$\sqrt{3}\sigma_{\rm a,p}$ through eq.~(2) in Biviano et
al. (2006). An estimate of the radius $\tilde{r_{\rm v}}$ is derived
through steps 7 and 9 in Biviano et al. (2006). Replacing the true
quantities $r_{\rm v}$ and $\sigma_{\rm a}$ with their estimates
$\tilde{r_{\rm v}}$ and $\sqrt{3}\sigma_{\rm a,p}$ in Fig.~4 in
Biviano et al. (2006), we obtain an estimate of $\sigma_{\rm v}$. The
cluster mass $M_{\sigma}$ is given by eq.~(2) in Biviano et al. (2006)
and is accurate to $\approx$10\%. The mass error is derived by
combining in quadrature the error in the velocity dispersion converted
in mass, additional velocity dispersion error introduced by the
uncertainty in Figure~4 in Biviano et al. (2006), and the 10\%
accuracy in $M_{\sigma}$. Typical mass errors are $\sim$20\%
(Table~\ref{t:sample}). To compute $r_{500}$ from $M_{200}$, we assume
an NFW (Navarro et al. 1997) model with the concentration parameter
given in step 7 in Biviano et al. (2006).

\begin{figure*}
\centering
\includegraphics[angle=270,width=8cm]{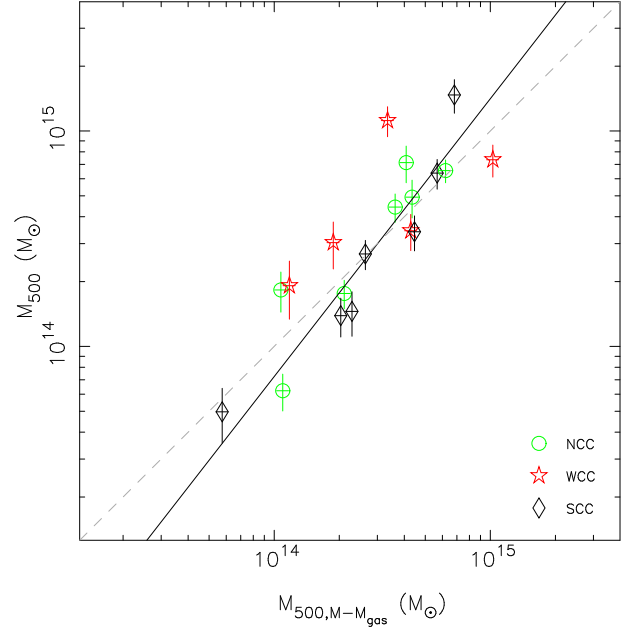}
\caption{Cluster mass derived from the velocity dispersion compared to
  the cluster mass derived from the gas mass using the mass vs. gas
  mass relation. The solid line shows the best fit. The dashed line
  denotes the 1:1 ratio. The green circles, red stars, and black
  diamonds stand for noncool-core, weak cool-core and strong cool-core
  clusters, respectively. The two most outliers are A2029 and A2065,
  which have the highest masses derived from the “harmonic” velocity
  dispersion.}
\label{f:m500}
\end{figure*}

As shown in Figure~3 in Zhang et al. (2011), the HIFLUGCS itself is
biased toward X-ray luminous clusters at the low-mass end and contains
an increasing fraction of mergers toward the high-mass end. This leads
to a large fraction of X-ray luminous clusters in the low-mass regime
and a dominant fraction of mergers in the high-mass regime for the
sample of 19 clusters. The velocity dispersion estimates can be biased
high by substructures, which leads to overestimating the dynamical
masses for mergers (e.g., Biviano et al. 2006). On the other hand, the
cluster masses, derived from our best fit of the X-ray hydrostatic
mass versus gas mass relation of the clusters and groups in Vikhlinin
et al. (2006), Arnaud et al. (2007), Pratt et al. (2009), and Sun et
al. (2009), are biased high for the X-ray luminous clusters in the
low-mass regime. Consequently, we observe that, at the
low-mass/high-mass end, the total masses derived from the velocity
dispersion appear low/high compared to the total masses derived from
the gas masses as shown in Figure~\ref{f:m500}. The two most outliers
are A2029 and A2065, which have the highest masses derived from the
``harmonic'' velocity dispersion. Therefore, the full sample of the 19
clusters and a sample of 17 clusters excluding A2029 and A2065 will be
investigated respectively, in Sect.~\ref{s:res}.

We derived all quantities consistently within the cluster
radius, $r_{500}$, unless explicitly stated otherwise.

\subsection{X-ray data}
\label{s:xray}

Details on the joint X-ray \emph{XMM-Newton} and \emph{ROSAT} data
analysis and properties of the \emph{XMM-Newton} observations can be
found in Zhang et al. (2009, 2011). The procedures for cleaning the
data are described in Sect. 2.2 and for detecting and subtracting
point-like sources in Sect.~2.3 in Zhang et al. (2009). The background
treatment can be found in Sect.~2.4 of Zhang et
al. (2009). Significant substructure features clearly detected in the
image are excised before we perform the spectral and surface
brightness analysis.

The truncation radii of the \emph{XMM-Newton} observed surface
brightness profiles, within which the signal-to-noise ratio (S/N) is
greater than 3, are rather small ($<r_{500}$). The \emph{ROSAT}
observed surface brightness profiles cover radii well beyond $r_{500}$
with S/N~$\ge 3$, although with sparse data points to resolve the
cluster core. As noted in Sect.~2.2 in Zhang et al. (2011), we
directly convert the \emph{ROSAT} surface brightness profile to the
\emph{XMM-Newton} count rate using the best-fit spectral model
obtained from the \emph{XMM-Newton} data. We then combine the
\emph{XMM-Newton} surface brightness profile within the truncation
radius, where the \emph{XMM-Newton} S/N is $\sim 3$, with the
\emph{ROSAT} converted surface brightness profile beyond the
truncation radius for further analysis.

The soft band (0.7--2~keV) X-ray surface brightness profile model
$S_{\rm X}(R)$, in which $R$ is the projected radius, is linked to the
ICM radial electron number density profile $n_{\rm e}(r)$ and
emissivity function as an integral performed along the line of sight
convolved with the \emph{XMM-Newton} point-spread function (PSF) as a
function of the detector coordinates $(X,Y)$,
\begin{equation}
  S_{\rm X}(R)\propto\int\int \int
  PSF(X,Y,R)\;   
  n^2_{\rm e}\left(\sqrt{l^2+R^2}\right) d\ell \;dX dY.
\label{e:sx}
\end{equation}
The gas mass distribution is derived by integrating the gas density
distribution. Gas masses are measured with respect to $r_{500}$
derived in Sect.~\ref{s:r500}. Errors are derived by combining in
quadrature measurement errors of gas masses at $r_{500}$ and errors
due to the variation of $r_{500}$ within its $1\sigma$
interval. Typical gas mass errors are within 10\% (see
Table~\ref{t:sample}).

The spectral analysis through which we obtain the radial temperature
distribution $T(r)$ is documented in Sect.~3 in Zhang et
al. (2009). The radial entropy distribution is computed as $S(r)=T(r)
n_{\rm e}^{-2/3}(r)$. The \emph{XMM-Newton} data enable measuring the
radial temperature distribution beyond $0.2r_{500}$. The central
entropies are thus measured at $0.1r_{500}$ and $0.2r_{500}$.

\subsection{Optical imaging data}

The SDSS\footnote{The $K$-band luminosity from the Two Micron All Sky
  Survey (2MASS) observations can also be used to measure the stellar
  mass (Lin et al. 2003). Compared to the 2MASS observations, the SDSS
  data allow us to make a more reliable selection of cluster member
  galaxies, which greatly reduces fore- and background
  contaminations.}  DR7 has the advantage over previous SDSS data
releases of using more reliable sky background subtraction (Abazajian
et al. 2009). We used the magnitudes in the ``$dered$'' table in the
``PHOTO'' catalog. Hereafter we briefly summarize the procedure on
constructing the GLF of the galaxy cluster (see Lagan\'a et al. 2008,
2011 for details) as follows.

A color-magnitude diagram is constructed for the galaxies within a
projected radius of $r_{500}$. The galaxies with $0.7 < (g-i) < 1.6$
and $i<18$ in the $g-i$ vs. $i$ diagram\footnote{This magnitude cut
  best suits the majority of the 19 clusters. For four clusters
  (IIIZw54, A1367, MKW4, and A400) we apply $i < 16$, which better
  defines the preliminary slope of the red sequence.} are bright and
red galaxies, and are used to make a preliminary fit of the red
sequence in the color-magnitude diagram with a linear
relation. Iteratively, the red sequence is fitted using those galaxies
in the full magnitude range within $3\sigma$ redward and blueward of
the previous red-sequence fit until the fit converges. The red
galaxies are defined as those within 0.3~mag redward and blueward in
$g-i$ of the best-fit red sequence, and blue galaxies as those bluer
than the lower limit for the red galaxy selection.

The fore- and background galaxies are subtracted statistically (e.g.,
Zwicky 1957; Oemler 1973). The $i$-band GLF is constructed for the
selected red and blue galaxies with a bin size of 0.5~mag. The GLF of
the background galaxies is derived using those galaxies within an
annular region located beyond $8r_{500}$ from the X-ray flux-weighted
cluster center. A power-law model is fitted to the GLF of the
background galaxies and subtracted from the overall GLF. A fair
background subtraction should not introduce major uncertainties since
the GLF is sampled well enough at the bright end, and it does not
abruptly rise at the faint end. Large-scale structures (LSSs) may
introduce some uncertainties into this kind of background
subtraction. Recent tests (e.g., Paolillo et al. 2001; Gonzalez et
al. 2007, Filippis et al. 2011) show that the difference in the GLF is
within 10\% compared to the case using the background level derived
from the SDSS GLF (e.g., Blanton et al. 2003), integrating over
redshift and applying mean evolutionary and $K$-corrections (e.g.,
Bruzual \& Charlot 2003) with the Padova isochrones (Bertelli et
al. 1994). Corrections due to poor or absent sampling of the faint end
of the GLF should only be a few per cent (see Andreon 2010).

There are two obvious components in the GLF for most clusters. A
double Schechter (1976) function is thus used to fit the GLF of each
cluster, in which the covariance between the power-law index at the
faint end ($\alpha$) and the characteristic magnitude ($M^{\ast}$) is
taken into account (see Table~\ref{t:glf}). The $K$-correction for
early-type galaxies and the redshift evolution correction given in
Poggianti (1997) are adopted to correct the characteristic
magnitudes. The total $i$-band luminosity is derived by integrating
the double Schechter function for magnitudes brighter than $-14$.

The mass-to-light ratios for ellipticals and spirals, respectively,
following Kauffmann et al. (2003) assuming the Salpeter (1955) initial
mass function (IMF) are adopted to compute the stellar mass from the
optical luminosity. The stellar mass estimate is tied to the choice of
IMF. Changing the IMF will scale the stellar mass estimate by a fixed
factor, e.g., from a Kroupa (2001) IMF to a Salpeter (1955) IMF with a
cut-off at $0.1{\rm M_{\odot}}$ results in a factor of 2 increase in
the stellar mass (e.g., Kauffmann et al. 2003).

Any systematic offset in the spectrophotometric calibration will
change the stellar mass estimate (e.g., Kauffmann et al. 2003). Fritz
et al. (2010) have carried out a thorough comparison of stellar mass
estimates for a large sample of more than 50 nearby clusters
($z<0.07$) computed from spectroscopic and photometric data. They
found that the stellar masses computed from the fiber-aperture
magnitude using SDSS DR7 photometric data are lower by $\sim 0.15$~dex
than the values computed from the spectroscopic data within the same
aperture (also see Andreon 2010). Our sample is in similar redshift
and mass ranges as their sample, and we also calculated the total
stellar mass using the SDSS DR7 data. Therefore, this bias is
corrected in our total stellar mass estimates.

\section{Results} 
\label{s:res}

\begin{figure*}
\centering
\includegraphics[angle=270,width=15cm]{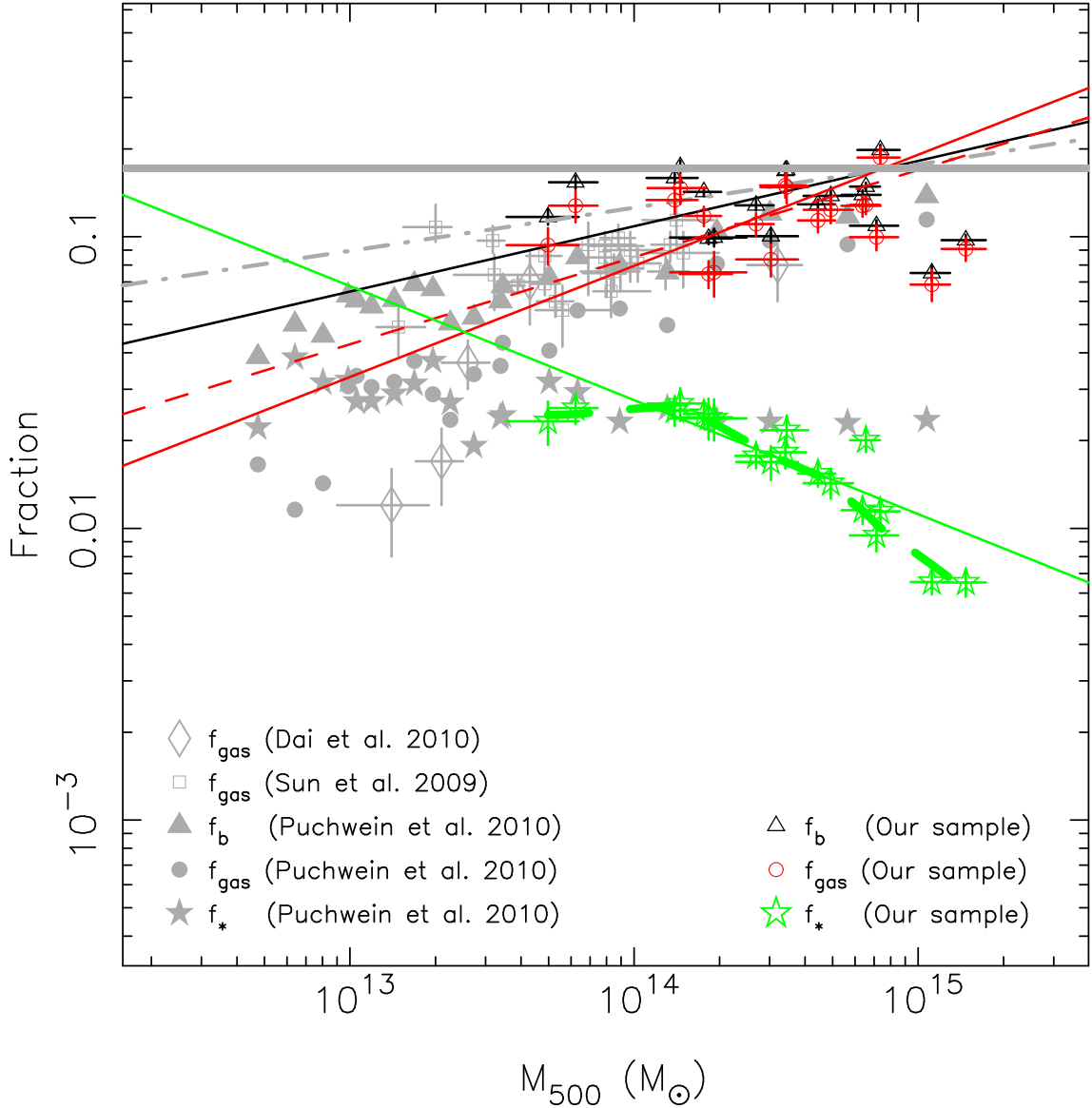}
\caption{Gas-mass fraction (red, open circles), stellar mass fraction
  (green, open stars), and baryon-mass fraction (black, open
  triangles) as a function of the total mass and the best fits
  excluding A2065 and A2029 in the same colors with solid line. The
  stellar mass fraction in our observational sample decreases with
  increasing cluster mass with (20$\pm$4)\% intrinsic scatter for the
  19 clusters. For comparison, we also show the gas-mass fraction
  (gray, solid circles), stellar mass fraction in galaxies (gray,
  solid stars), and the sum of these two fractions (gray, solid
  triangles) for the simulated sample of 21 clusters with AGN feedback
  in Puchwein et al. (2010), as well as the gas mass fractions of the
  X-ray selected groups in Sun et al. (2009; gray, open boxes) and
  near-infrared selected groups in Dai et al. (2010; gray, open
  diamonds).  The best fit of the baryon-mass fraction as a function
  of the total mass of the observational sample in Lin et al. (2003)
  is shown in gray with dot-dashed line. The gray band shows the
  $1\sigma$ measurement from the WMAP 5-year result (Dunkley et
  al. 2009). A2142 displays a baryon-mass fraction of $0.198\pm
  0.008$, which exceeds the WMAP result with a $3\sigma$
  significance. The best fit of the gas-mass fractions combining our
  clusters and the clusters in Sun et al. (2009) is shown in red with
  dashed line with (26$\pm$8)\% intrinsic scatter. }
\label{f:frac}
\end{figure*}

Figure~\ref{f:frac} shows the gas-mass, stellar mass, and baryon-mass
fractions as a function of the total mass. We also compared our
observational sample to some existing observational samples (e.g., Lin
et al. 2003; Gonzalez et al. 2007; Giodini et al. 2009; Pratt et
al. 2009; Sun et al. 2009; Andreon 2010; Dai et al. 2010), as well as
simulated samples (e.g., Puchwein et al. 2008, 2010; Fabjan et
al. 2010).

\subsection{Gas-mass fraction}

The gas-mass fractions of the 19 clusters appear flat in the given
narrow mass range with $(26\pm4)$\% intrinsic scatter. Excluding the
two most prominent outliers, A2029 and A2065, shown in
Sect.~\ref{s:r500}, the gas-mass fraction increases with the total
mass as $f_{\rm gas,500}=10^{-(1.10\pm0.16)} ( M_{500}/[10^{14}{\rm
  M_{\odot}}] )^{0.38\pm 0.36}$ with $(23\pm17)$\% intrinsic
scatter. To constrain the slope precisely, we combine our 19 clusters
and the groups in Sun et al. (2009), which show an increasing gas-mass
fraction as a function of the total mass, i.e., $f_{\rm
  gas,500}=10^{-(1.07\pm0.02)} ( M_{500}/[10^{14}{\rm M_{\odot}}]
)^{0.30\pm 0.07}$ with $(26\pm 8)$\% intrinsic scatter. This shows the
importance of having a wide mass range to calibrate the mass
dependence of the gas-mass fraction.

The normalization for our sample is comparable to what is obtained for
the X-ray selected cluster sample in Pratt et al. (2009) and X-ray
selected group sample in Sun et al. (2009), and is higher than the
2MASS selected stacked groups in Dai et al. (2010) with $>3\sigma$
significance. This indicates that to some extent X-ray selected groups
may be biased toward systems with high gas-mass fractions, whereas
near-infrared (or optically) selected groups are biased toward systems
with low gas-mass fractions. (The extreme systems are probably the
so-called X-ray under-luminous systems, e.g., Dietrich et al. 2009.)

The gas-mass fractions of our observational sample agree well with the
predictions of the simulated sample, in particular, in the high-mass
regime, i.e., $\ge 10^{14}{\rm M_{\odot}}$. It is interesting that the
slope of the mock sample with AGN feedback is between the measured
slopes of X-ray selected and near-infrared selected samples. This can
either be because that the simulation sample is free of selection
biases, or our sample has a sampling bias in the low-mass regime
caused by low-quality statistics.

\subsection{Stellar mass fraction}

The stellar mass fraction in our observational sample decreases with
increasing cluster mass; i.e., $f_{\rm *,500}=10^{-(1.53\pm0.05)}(
{M_{500}}/[{10^{14}{\rm M_{\odot}}}] )^{-(0.49\pm 0.09)}$ with $(20\pm
4)$\% intrinsic scatter for the 19 clusters. Excluding A2029 and
A2065, the best fit is $f_{\rm *,500}=10^{-(1.56\pm0.05)}(
{M_{500}}/[{10^{14}{\rm M_{\odot}}}] )^{-(0.39\pm 0.09)}$ with $(15\pm
29)$\% intrinsic scatter. We also derived the local regression fit
used in Zhang et al. (2009), which demonstrates the flattening of the
stellar mass fraction in the mass range lower than
$2\times10^{14}M_{\odot}$ (see Fig.~\ref{f:frac}). There are three
situations that may introduce measurement uncertainties and affect the
best fit, (1) the poor sampling in the low-mass regime, i.e.,
$<10^{14}{\rm M_{\odot}}$, (2) the potential volume dependence in the
background estimate, and (3) the two outliers for which the total
masses are likely overestimated. The volume fraction of the background
galaxies decreases with decreasing cluster redshift. The estimate of
the local background in deriving the GLF becomes less reliable in,
e.g., Coma ($r_{500}=41.1^{\prime}$).

Within $1 \sigma$, the best fit of the stellar mass fraction as a
function of the total mass agrees between our sample and the published
samples (e.g., Lin et al. 2003; Gonzalez et al. 2007; Giodini et
al. 2009; Andreon 2010; Dai et al. 2010) as shown in
Table~\ref{t:fit}. The stellar mass fractions of nearly half of the 19
clusters in the observational sample agree with those of the simulated
samples within their $1\sigma$ interval (e.g., Fabjan et al. 2010;
Puchwein et al. 2010). The amount of the stellar mass in galaxies in
simulations depends rather sensitively on how the supernova feedback
is included (e.g., Springel \& Hernquist 2003a; Borgani et
al. 2006). For example, using kinetic feedback, e.g., in addition to
the thermal feedback that Puchwein et al. (2010) used, one can obtain
up to a factor of 2 fewer stars. Furthermore, the cluster mass is well
defined in simulations, while there are known uncertainties in mass
measurements of galaxy clusters (e.g., Biviano et al. 2006). This also
causes different trends and scatter between the observational and
simulated samples.

\subsection{Baryon-mass fraction}

In simulations, part of the gas does not enter the potential well of a
galaxy group because gas is collisional. This phenomenon drives the
so-called gas depletion. This ``correction'' is usually taken into
account when estimating the baryon-mass fractions for observed
groups. It amounts to roughly 10\% within $r_{500}$ for groups (e.g.,
Frenk et al. 1999; Kay et al. 2004). The gas depletion is only a few
percent in galaxy clusters as massive as those considered in this
study (e.g., Ettori et al. 2006; Fabjan et al. 2010). We therefore do
not account for it, and define the baryon-mass fraction of the galaxy
cluster as the sum of the stellar mass fraction and gas-mass fraction,
$f_{\rm b,500}:=M_{\ast,500}/M_{500} + M_{\rm gas,500}/M_{500}$.

In Fig.~\ref{f:frac}, we also show the baryon-mass fraction as a
function of the total mass. The baryon-mass fractions of all 19
clusters appear flat as a function of the total mass. Even when A2029
and A2065 are excluded, the best fit, $f_{\rm b,500}=10^{-(0.964\pm
  0.250)} ( {M_{500}}/[{10^{14}{\rm M_{\odot}}}] )^{0.224\pm 0.567}$,
still appears flat as a function of the total mass within its
$1\sigma$ error.

As shown in Table~\ref{t:fit}, the baryon-mass fractions of our sample
display a similar trend as that exhibited by other observational
samples (e.g., Lin et al. 2003; Gonzalez et al. 2007; Giodini et
al. 2009; Andreon 2010). According to the best fit for our sample, the
value reaches the WMAP 5-year measurement of the cosmic baryon-mass
fraction (Dunkley et al. 2009) at $\sim 7.7\times 10^{14}{\rm
  M_{\odot}}$. The observational sample displays large scatter
suffering from measurement uncertainties and probably also cluster
physics.

\section{Discussions}
\label{s:dis}

\subsection{ICL}

The small discrepancy between the baryon-mass fraction of our
observational sample and the cosmic value appears to increase toward
the low-mass end. The ICL is suggested to be one of the most important
forms of missing baryons that accounts for this discrepancy, and
accounts for 6\%--22\% of the total cluster light in the $r$-band
(e.g., Krick \& Bernstein 2007; Gonzalez et al. 2007; Zibetti et
al. 2005). Pierini et al. (2008) suggest that the ICL fraction may
increase in cluster mergers.

Simulations with different cluster physics (e.g., Frenk et al. 1999;
Borgani et al. 2004; Kay et al. 2004; Kravtsov et al. 2005; Nagai et
al. 2007; Evrard et al. 2008; Sijacki et al. 2007; Fabjan et al. 2010;
Kapferer et al. 2010; Puchwein et al. 2010) give different
predictions. Kravtsov et al. (2005) performed numerical simulations
that assume nonradiative hydrodynamics (without dissipations) in one
run and radiative cooling, as well as several other physical processes
in the other run. They find that the baryon-mass fraction does not
depend on the cluster mass in nonradiative simulations, and that the
predicted value within $r_{500}$ is 5\% below the cosmic value. In
simulations with radiative cooling, star formation, and AGN feedback,
both the baryon-mass fraction and the gas-mass fraction increase with
increasing mass (e.g., Kravtsov et al. 2005; Puchwein et al. 2010). In
the runs with AGN feedback, the baryon-mass fraction is close to the
cosmic value for the most massive clusters, but is significantly lower
than the cosmic value for low-mass systems (e.g, Fabjan et al. 2010;
Puchwein et al. 2010). This is due to the amount of the ICL, as well
as of a significant amount of gas removed by AGN heating from the
central regions of clusters and driven to $>r_{500}$, in particular,
in low-mass systems as shown by the sum of the baryon-mass fraction
($f_{\rm gas,500}+f_{\ast,500}$) and the ICL in Figure~5 in Puchwein
et al. (2010).

In Puchwein et al. (2010), the ICL amounts to $\sim 58$\% of the total
stars, i.e., the sum of the ICL and stars in cluster galaxies, in a
cluster of $4\times 10^{13}{\rm M_{\odot}}$. The amount of the stellar
mass in the ICL found in simulations is enough to explain the
difference between the baryon-mass fraction of our observational
sample and the WMAP 5-year predicted cosmic value in the low-mass
regime. However, such a large amount of the stellar mass in the ICL
seems to be incompatible with recent observations (e.g., Krick \&
Bernstein 2007; Gonzalez et al. 2007; Zibetti et al. 2005). This
suggests that the ICL only accounts for part of the missing baryons
and that other mechanisms like gas expulsion by AGN heating may be
important.

\subsection{Star-formation efficiency}

We observe a decreasing stellar mass fraction as a function of the
total mass (Fig.~\ref{f:frac}). The stellar mass fraction has been
widely used as an estimate of the star-formation
efficiency\footnote{The star-formation efficiency used in this work is
  not the stellar-to-initial molecular hydrogen mass ratio, but the
  stellar-to-total cluster mass ratio.} (e.g., Bryan 2000). There are
no correlations between the stellar-to-gas mass ratio within $r_{500}$
and redshift. This indicates that the mass dependence of the
star-formation efficiency is not a selection effect of redshift
evolution.

Other observational samples also show varying star-formation
efficiency with total mass (e.g., Gonzalez et al. 2000; Lin et
al. 2003; Lagan\'a et al. 2008, 2011; Ettori et al. 2009). Dai et
al. (2010) suggest that the fraction of baryon loss is determined by
the depth of the potential well of the system, which leads to an
increasing baryon-mass fraction as a function of the increasing total
mass of a cluster. Toward the high-mass end, it approaches the cosmic
value. Our finding supports two interpretations demonstrated in
simulations (e.g., Fabjan et al. 2010). In massive clusters, more hot
gas and dark matter is settled in the deep potential wells than
individual galaxies. In low-mass systems, the accretion of individual
less massive galaxies is more important, and more low entropy gas is
brought in to form stars. Both processes result in low star-formation
efficiency in massive systems.

\subsubsection{Mergers and energy inputs from feedback}

In the hierarchical structure-formation scenario, the efficiency of
stripping gas from member galaxies depends on the total mass of the
system. In more massive clusters, galaxies exhibit higher velocity
dispersion due to their deep potential wells, and as a consequence the
efficiency of stripping gas from member galaxies is higher, which
diminishes the star-formation efficiency in the galaxies. In low-mass
systems, heating due to merging is less efficient because of their
shallow potential wells. On the other hand, the effect of AGN feedback
is significant in groups due to their shallow potential wells. Giodini
et al. (2010) find that feedback from radio-mode AGN heating can
account for the reduced gas fractions in groups in the COSMOS
field. Simulations indeed show that a significant amount of gas has
been removed by AGN heating from the central regions and been driven
to the region beyond $r_{500}$ (e.g., Fabjan et al. 2010; Puchwein et
al. 2010).  Furthermore, the fraction of X-ray luminous clusters is
high in the low-mass regime because of the nature of the flux
selection of the HIFLUGCS. Most X-ray luminous systems are cool-core
clusters and they host AGN in the central regions.  There may be
correlations between the stellar-to-gas mass ratio and the indicators
of merging and various energy inputs from feedback.

\begin{figure*}
\centering
\includegraphics[angle=270,width=8cm]{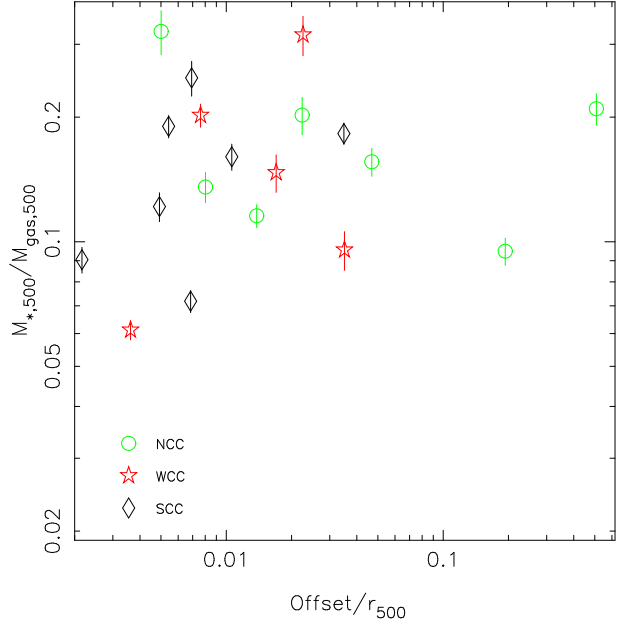}
\caption{Stellar-to-gas mass ratio within $r_{500}$ vs. offset between
  the X-ray flux-weighted center and the BCG position. The symbols have 
the same meaning as those in Fig.~\ref{f:m500}. The intrinsic scatter 
is (41$\pm$8)\% and (66$\pm$11)\% for the ten high- and nine
low-offset systems, respectively. This indicates that the scatter
of the stellar-to-gas mass ratios may reflect the freedom in the
formation history of the systems since the last major merging
episode.
}
\label{f:msmg_deltamag}
\end{figure*}

As shown in Fig.~\ref{f:msmg_deltamag}, the scatter in the
stellar-to-gas mass ratios appears to increase with decreasing offset
between the X-ray flux-weighted center and the bright cluster galaxy
(BCG). We divided the sample into two comparable-size subsamples with
a threshold of $0.01r_{500}$ offset, and found $(41\pm 8)$\% and
$(66\pm11)$\% intrinsic scatter for the ten high- and nine low-offset
systems, respectively. This indicates that the scatter of the
stellar-to-gas mass ratios may reflect the freedom in the formation
history of the systems since the last major merging episode.

Neither the stellar-to-gas mass ratio nor the offset between the X-ray
flux-weighted center and the BCG correlates with the $i$-band
magnitude difference between the BCG and the second brightest cluster
galaxy. There are several possible explanations. The epoch of major
star formation in the progenitors of massive ellipticals may happen
much earlier than they fell into the cluster potential well. There may
be smoother accretion of dark matter and hot gas than individual
galaxies in quiescent systems. There may be a large fraction of the
stellar mass in the ICL since mergers and tidal stresses have already
happened and dynamical friction had more time to operate before the
time of observations.

There is no correlation between the stellar-to-gas mass ratio and the
BCG magnitude. It indicates that the bulk of star formation does not
dominate in the central region of the cluster, where the BCG is (also
see Edwards et al. 2007; Loubser et al. 2009; Wang et al. 2010). The
star-formation efficiency may be similar in the cluster core and the
outskirts.

From the X-ray data we derived the cooling radius, $r_{\rm cool}$,
within which the cooling time is shorter than the age of the
Universe. There is no evidence of high star-formation efficiency with
increasing X-ray luminosity within the cooling radius. This indicates
that radiative cooling losses within the cooling radii are offset by
some heating mechanism(s).

\begin{figure*}
\centering
\includegraphics[angle=270,width=8cm]{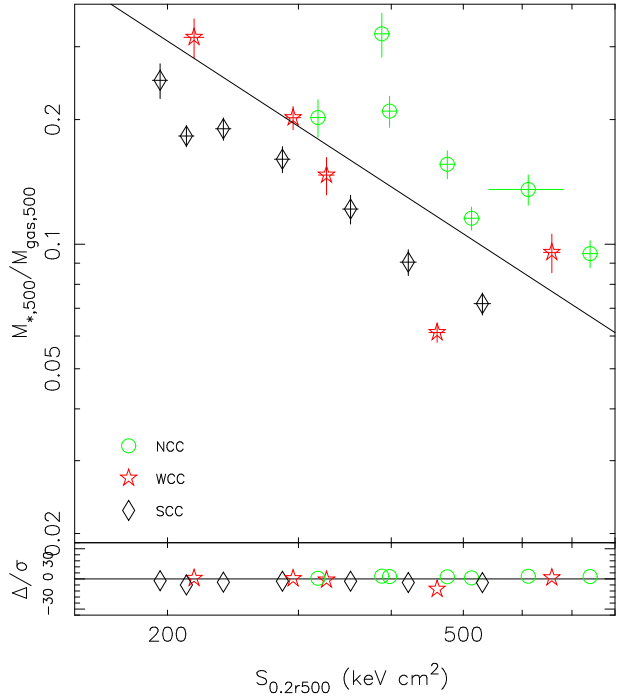}
\caption{Stellar-to-gas mass ratio within $r_{500}$ vs. gas entropy at
  $0.2r_{500}$ and the best fit in the upper panel, as well as the
  residuals normalized by the error bars in the lower panel. The
  symbols have the same meaning as those in Fig.~\ref{f:m500}. The
  data support the interpretation that heating from merging quenches
  the star-formation activity of galaxies in massive systems, and
  feedback from supernovae and/or radio galaxies drives a significant
  amount of gas to the regions beyond r500 or, alternatively, a
  substantially higher stellar mass fraction in the ICL is present in
  nonrelaxed systems (e.g., Pierini et al. 2008).}
\label{f:msmg_en}
\end{figure*}

The stellar-to-gas mass ratio decreases with increasing entropy at
$0.1r_{500}$ and $0.2r_{500}$ (Fig.~\ref{f:msmg_en}), i.e.,
${M_{\ast,500}}/{M_{\rm gas,500}}=10^{-(0.477\pm0.075)}
{S_{0.1r_{500}}}/[{100{\rm {keV\;cm}^2}}] )^{-(1.03 \pm 0.20)}$ with
$(51\pm 7)$\% intrinsic scatter and ${M_{\ast,500}}/{M_{\rm
    gas,500}}=10^{-(0.157\pm 0.113)} ({S_{0.2r_{500}}}/[{100{\rm
    {keV\;cm}^2}}] )^{-(1.17 \pm 0.19)}$ with $(41\pm 5)$\% intrinsic
scatter. For the cool-core clusters alone, the scatter of the
stellar-to-gas mass ratios becomes smaller, e.g.,
${M_{\ast,500}}/{M_{\rm gas,500}}=10^{(2.18\pm0.49)}
({S_{0.1r_{500}}}/[{100{\rm {keV\;cm}^2}}] )^{-(1.17 \pm 0.19)}$ with
$(41\pm 5)$\% intrinsic scatter. The data support the interpretation
that heating from merging quenches the star-formation activity of
galaxies in massive systems, and feedback from supernovae and/or radio
galaxies drives a significant amount of gas to the regions beyond
$r_{500}$ or, alternatively, a substantially higher stellar mass
fraction in the ICL is present in nonrelaxed systems (e.g., Pierini
et al. 2008).

Given the evidence of AGN feedback with radio-jet powered energy input
in nearby clusters (e.g., Birzan et al. 2008), we calculated the
cavity power following the correlation in Birzan et al. (2008) from
the radio bolometric luminosity given in Mittal et al. (2009).
Unfortunately, we found no correlation between the stellar-to-gas mass
ratio and the ratio between the X-ray bolometric luminosity within the
cooling radius and the cavity power, respectively. The data do not
support our guess on heating from the ongoing AGN feedback.

The explanation may be that the cavity power provided by recent AGN
activities only affects the current star-formation rate and it has
less effect on more massive systems, in which energy input from
merging becomes more important. The stellar-to-gas mass ratio is set
by the whole star-formation and energy-input histories of the galaxy
clusters, and not by the current star-formation rates of the member
galaxies.

\subsubsection{Metal enrichment}

The iron abundance of the ICM cannot be measured in a homogeneous
radial range for nearby clusters due to the small field-of-view (FOV)
of \emph{XMM-Newton} compared to the sizes of the clusters. This
causes scatter in such measurements. On one hand, \emph{XMM-Newton}
FOV covers more of the cluster areas for lower-mass systems for a
fixed redshift. Since the universal iron abundance decreases with
increasing cluster radius (e.g., Kapferer et al. 2009, 2010), the
measured global iron abundance within the \emph{XMM-Newton} FOV is
thus biased lower for low-mass systems than for high-mass systems. On
the other hand, the X-ray emission of less massive systems is traced
out to smaller fractions of $r_{500}$ than for more massive systems
according to the flux limit. This should result in higher metallicity
measurements in less massive systems when all systems can be observed
out to large radii, e.g., $\sim 0.5 r_{500}$. We normalized the
radius, within which the global temperature and metallicity are
measured, by $r_{500}$, and found that the normalized value has no
correlation with $r_{500}$. Therefore both effects are negligible for
our sample.

\begin{figure*}
\centering
\includegraphics[angle=270,width=8cm]{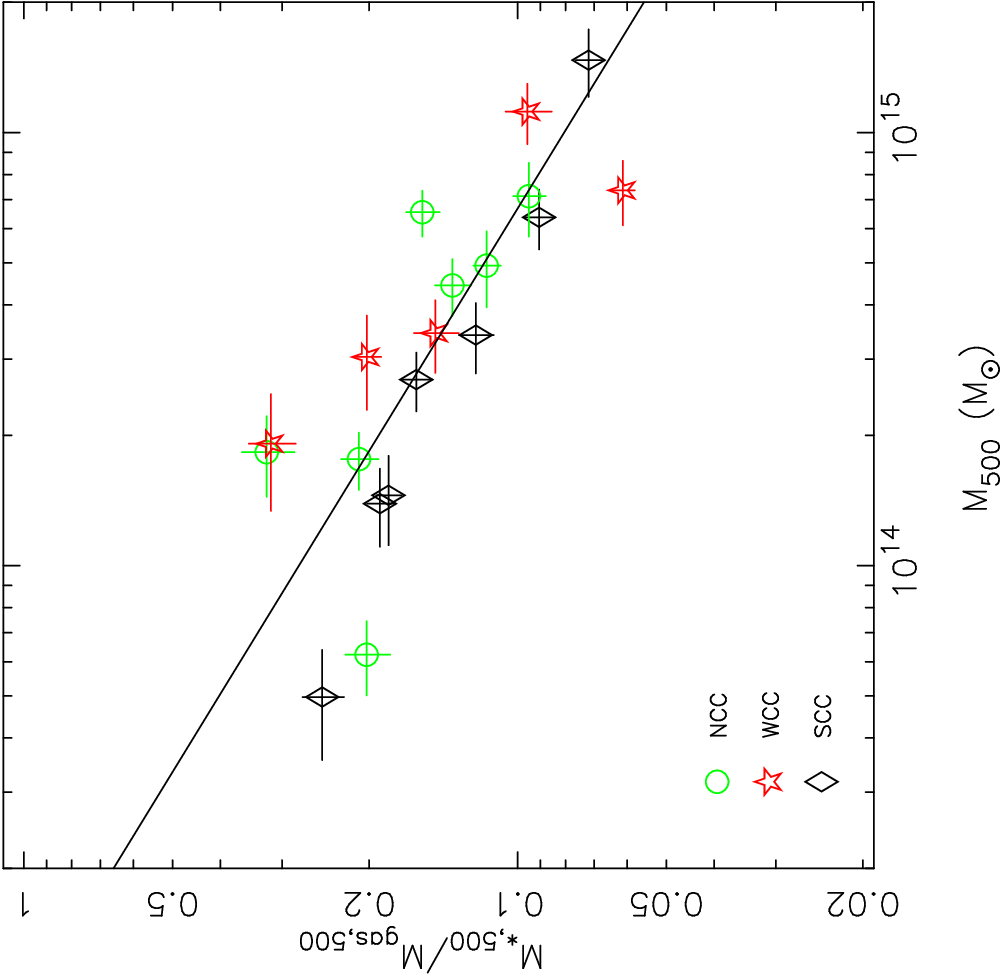}
\includegraphics[angle=270,width=8cm]{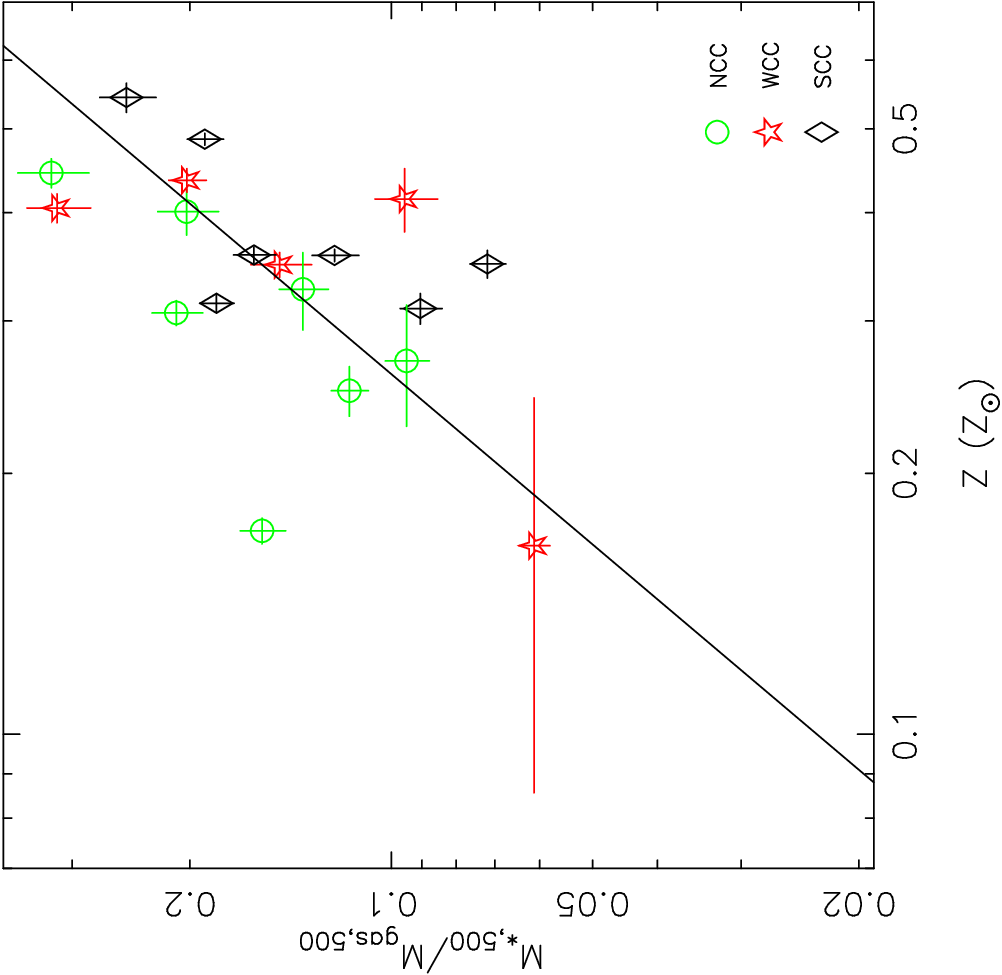}
\caption{Stellar-to-gas mass ratio within $r_{500}$ vs. total mass
  (left panel) and iron abundance (right panel) and their best
  fits. The symbols have the same meaning as those in
  Fig.~\ref{f:m500}. These two correlations indicate that the iron in
  the ICM mainly comes from the pollution by the star formation that
  happened in the past. In less massive galaxy systems, the star
  formation efficiency is higher.}
\label{f:msmg_abun}
\end{figure*}

As shown in the left hand panel in Fig.~\ref{f:msmg_abun}, we found an
increasing stellar-to-gas mass ratio with decreasing cluster mass,
i.e., ${M_{\ast,500}}/{M_{\rm gas,500}}= 10^{-(0.557\pm 0.071)}
({M_{500}}/[{10^{14}{\rm M_{\odot}}}])^{-(0.537\pm 0.101)}$ with
$(29\pm5)$\% intrinsic scatter. The cool-core (classified as ``S'' in
Table~\ref{t:sample}) clusters display a flatter slope, $-(0.417\pm
0.067)$, and smaller intrinsic scatter, $<17$\%, than the noncool-core
(classified as ``W'' and ``N'' in Table~\ref{t:sample}) clusters,
which show a slope of $-(0.617\pm 0.228)$ and $(38\pm20)$\%
scatter. There is an increasing stellar-to-gas mass ratio with
increasing iron abundance of the ICM in the right hand panel in
Fig.~\ref{f:msmg_abun}; i.e., ${M_{\ast,500}}/{M_{\rm gas,500}}=
10^{-(0.106 \pm 0.174)} ({Z}/{\rm Z_{\odot}} )^{1.53 \pm 0.38}$ with
$(44\pm6)$\% intrinsic scatter. We note that the trend holds even when
one looks at the cool-core clusters and noncool-core clusters,
respectively. It is interesting that the scatter of the cool-core
clusters and noncool-core clusters is comparable.

The above two correlations indicate that the iron in the ICM mainly
comes from the pollution by the star formation that happened in the
past. In less massive galaxy systems, the star formation efficiency is
higher. In other words, more stars were formed that have delivered
more metals to enrich the hot gas. In high-mass clusters, energy
feedback from, e.g., merging, may quench star formation in their
member galaxies, which results in low star-formation efficiency and
less metal enrichment in the hot gas by stars. In addition, a larger
amount of hot gas is accreted in a more massive system due to its
deeper potential well, which dilutes the iron abundance more.

\begin{figure*}
\centering
\includegraphics[angle=270,width=8cm]{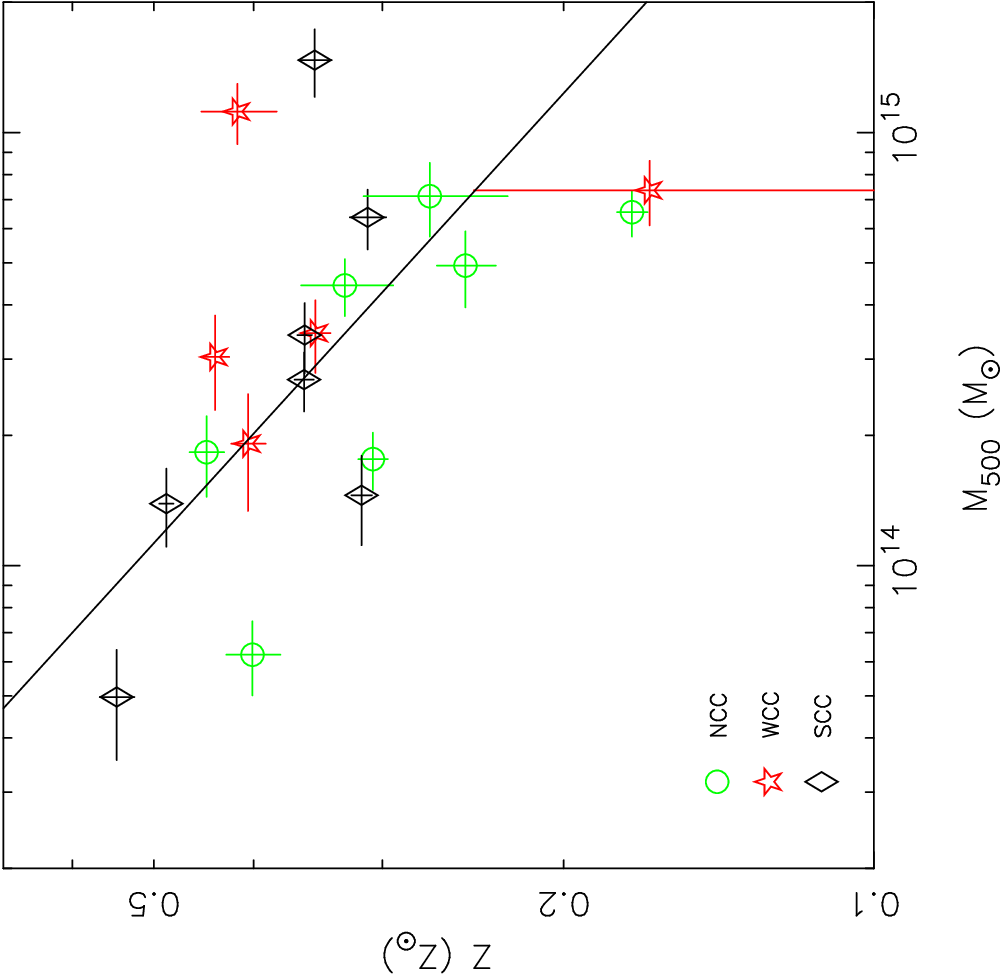}
\includegraphics[angle=270,width=8cm]{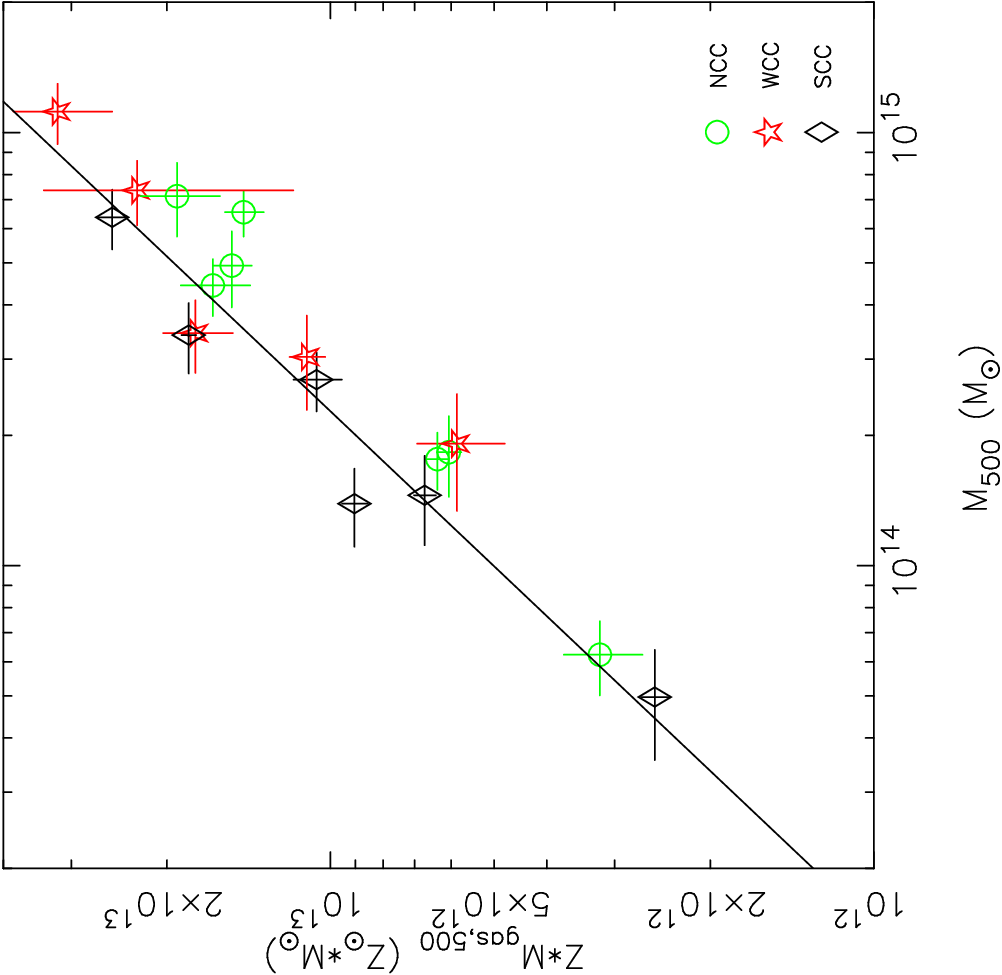}
\caption{Iron abundance (left panel) and iron mass (right panel)
  vs. total mass and their best fits. The symbols have the same
  meaning as those in Fig.~\ref{f:m500}. This finding agrees with the
  trend seen in simulations (e.g., Figure 11 in Fabjan et al. 2010)
  toward less massive clusters having lower gas-mass fractions but
  higher iron-mass fractions, so that the gas is more metal-rich in
  those systems.}
\label{f:abun_m500}
\end{figure*}

\begin{figure*}
\centering
\includegraphics[angle=270,width=8cm]{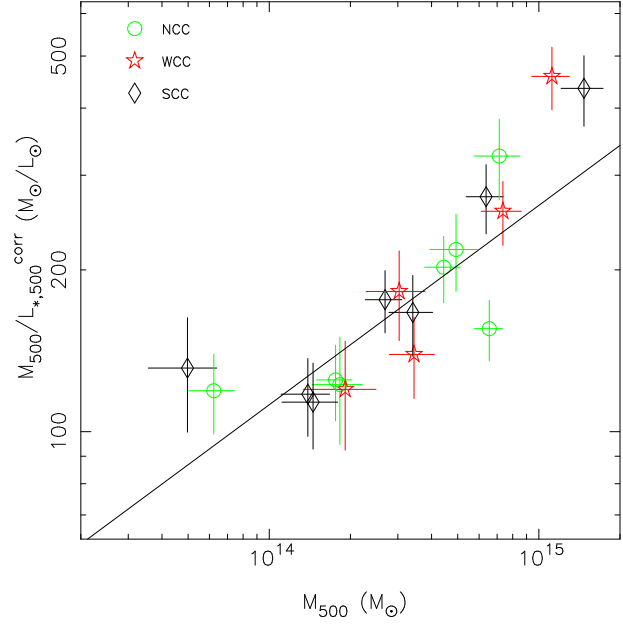}
\caption{Total mass-to-optical light ratio within $r_{500}$ vs. total
  mass and the best fit excluding A2029 and A2065. The symbols have
  the same meaning as those in Fig.~\ref{f:m500}. The behavior of the
  total mass-to-optical light ratio supports the variation in
  star-formation efficiencies.}
\label{f:mlr_m500}
\end{figure*}

As shown in Fig.~\ref{f:abun_m500}, the observational sample shows
decreasing iron abundance with increasing total cluster mass, i.e.,
${Z}/{\rm Z_{\odot}}=10^{-(0.28\pm 0.07)} ({M_{500}}/[{10^{14}{\rm
    M_{\odot}}}] )^{-(0.383\pm 0.101)}$ with $(38\pm 5)$\% intrinsic
scatter, and increasing iron mass with increasing total cluster mass,
i.e., ${Z M_{\rm gas,500}}/[{\rm Z_{\odot}M_{\odot}}]=10^{12.7\pm
  0.03} ({M_{500}}/[{10^{14}{\rm M_{\odot}}}] )^{0.842\pm 0.049}$ with
$(17\pm 3)$\% intrinsic scatter.  This finding agrees with the trend
seen in simulations (e.g., Figure~11 in Fabjan et al. 2010) that less
massive clusters have lower gas-mass fractions but higher iron-mass
fractions, so that the gas is more metal-rich in those systems. This
can be explained in two ways. One is that the star-formation
efficiency is high in low-mass systems: more gas is converted into
stars within $r_{500}$ which yields more metals that enrich the
gas. Alternatively, metal-poor hot gas has not fallen into the
potential wells in low-mass systems. Most likely, metal-loaded
outflows from galaxies started well before the potential well of a
cluster was settled (see Noll et al. 2009), so that the metal
enrichment of the observed ICM reflects a heavier star-formation
activity in low-mass systems. Accordingly, the total mass-to-optical
light ratio decreases with decreasing cluster mass
(Fig.~\ref{f:mlr_m500}).

\subsubsection{Total mass-to-optical light ratio}

Optical and near-infrared observations suggest that the total
mass-to-optical light ratio, i.e., $M/L_{\ast}^{\rm corr}$, increases
with cluster mass with a slope in the range of 0.2--0.4 assuming a
power-law relation (e.g., Adami et al. 1998; Bahcall \& Comerford
2002; Rines et al. 2004). In Fig.~\ref{f:mlr_m500}, we show the total
mass-to-optical light ratio as a function of the total mass for our
sample, which follows a trend of $M_{500}/L_{\ast,500}^{\rm corr} =
10^{2.02\pm 0.05} ( {M_{500}}/[{10^{14}{\rm M_{\odot}}}] )^{0.468\pm
  0.085}$ with $(17\pm 4)$\% intrinsic scatter. When A2029 and A2065
are excluded, the best fit is $M_{500}/L_{\ast,500}^{\rm corr} =
10^{2.05\pm 0.05} ( {M_{500}}/[{10^{14}{\rm M_{\odot}}}] )^{0.371\pm
  0.098}$ with $(12\pm 10)$\% intrinsic scatter.

The behavior of the total mass-to-optical light ratio supports the
variation in star-formation efficiencies found in simulations (e.g.,
Springel \& Hernquist 2003a; Saro et al. 2006).  More massive clusters
host more massive galaxies dominated by old, passively evolving
stellar populations, where optical light has declined more
significantly than the average coeval galaxy due to their earlier
formation time. Alternatively, more massive clusters may have accreted
many more low-mass galaxies, or inhibited star formation in
intermediate-mass/low-mass galaxies in a more efficient way, which
leads to higher total mass-to-optical light ratios in massive systems.

The scatter in the total mass-to-optical light ratio vs. total mass
may indicate the coupling between the stochastic nature of the
assembly times of dark matter halo and stellar mass. The best fit
excluding A2029 and A2065 leads to $M_{500}/L_{\ast,500}^{\rm corr}
\sim 264$ for a system of $10^{15}{\rm M_{\odot}}$. Given the galaxy
luminosity density and the assumption of an equivalent total
mass-to-optical light ratio of the Universe to our extrapolated value
for a $M_{500}=10^{15}{\rm M_{\odot}}$ system, this results in a
consistent $\Omega_{\rm m}$ with the concordance value within the
scatter (also see e.g., Girardi et al. 2000; Reiprich 2003).

\section{Conclusions} 
\label{s:conc}

We have investigated the baryon content of a sample of 19 clusters
selected from an X-ray flux-limited sample of 64 nearby clusters, in
which we measured the cluster total masses and radii based on the
optical spectroscopic data, which are independent of both the X-ray
\emph{XMM-Newton} and \emph{ROSAT} data that we used to derive X-ray
quantities, such as the gas masses, and of the SDSS DR7 imaging data
that we used to compute the stellar masses. We summarize our results
and interpretations as follows.
 
\begin{itemize}

\item The gas-mass fraction of the observational sample increases with
  increasing cluster mass. The observed gas-mass fractions of our
  sample agree with previous observational results (e.g., Sun et
  al. 2007; Pratt et al. 2009). The values of our observational sample
  are also in good agreement with those of the simulated samples
  (e.g., Fabjan et al. 2010; Puchwein et al. 2010) in the high-mass
  regime.

\item The stellar mass fraction decreases with increasing cluster mass
  and exhibits large scatter ($20\pm 4$\%). The observed stellar mass
  fractions of our sample agree with previous observational results
  and the results of the simulated clusters (e.g., Fabjan et al. 2010;
  Puchwein et al. 2010) within the scatter.

\item We observe lower stellar-to-gas mass ratios in those systems
  that display higher central entropies indicating energy inputs from
  feedback. This supports a twofold interpretation: heating from
  merging quenches the star formation in galaxies in massive systems,
  and feedback from supernovae and/or radio galaxies removes a
  significant amount of gas in low-mass systems or, alternatively, a
  substantially higher stellar mass fraction in the ICL is
  associated with dynamically active systems of galaxies.

\item Less massive clusters are confirmed to host less gas per unit
  total mass; however, they exhibit higher metal-mass fractions, so
  that their ICM is more metal-rich. Member galaxies of low-mass
  systems may be forming stars more efficiently, thereby producing
  more metals, part of which are ejected from the galaxies and
  enriched the ICM to a higher level. Alternatively, some gas could
  have been expelled from these systems or be missing from the hot
  phase. The former hypothesis is preferred as the main driver of the
  mass-dependent metal enrichment since the total mass-to-optical
  luminosity ratio increases as the total mass increases.

\end{itemize}

\bigskip 

\begin{acknowledgements}

  The \emph{XMM-Newton} project is an ESA Science Mission with
  instruments and contributions directly funded by ESA Member States
  and the USA (NASA). The \emph{XMM-Newton} project is supported by
  the Bundesministerium f\"ur Wirtschaft und Technologie/Deutsches
  Zentrum f\"ur Luft- und Raumfahrt (BMWI/DLR, FKZ 50 OX 0001) and the
  Max-Planck Society. Part of the simulations, to which our results
  were compared, were performed by Debora Sijacki on the Cambridge
  high-performance computing cluster Darwin. Y.Y.Z. acknowledges Pavel
  Kroupa and Stefano Borgani for constructive discussions and Jacopo
  Fritz for providing the best-fit relation between stellar masses
  computed from the SDSS DR7 photometric data and spectroscopic
  data. Y.Y.Z. acknowledges support from the German BMBF through the
  Verbundforschung under grant No.\,50\,OR\,1005 and travel support
  from the Deutsche Forschungsgemeinschaft Priority Program 1177
  (Witnesses of Cosmic History: Formation and Evolution of Galaxies,
  Black Holes and Their Environment). T.F.L acknowledges support from
  the FAPESP through grants 2006/56213-9 and 2008/04318-7, as well as
  from the CAPES through BEX3405-10-9. D.P. acknowledges support from
  the German BMBF through the Verbundforschung under grant
  No.\,50\,OR\,0405 and the kind hospitality of the MPE. E.P. is
  supported by the DFG through Transregio 33. T.H.R. acknowledges
  support by the DFG through Heisenberg grant RE 1462/5.

\end{acknowledgements}

\bibliography{Abell}

\clearpage

\begin{center}
\begin{table*}
  \caption{Properties of the 19 galaxy clusters}
\begin{tabular}{ccccccccc}
  \hline 
  Name & \multicolumn{2}{c}{X-ray center (J2000)} & redshift & $M_{500}$ & $M_{\rm 500,M-M_{gas}}$ & $M_{gas,500}$ & $M_{*,500}$   & Undisturbed \\

  & R.A. & DEC &   & $10^{14} {\rm M_{\odot}}$ & $10^{14} {\rm M_{\odot}}$ & $10^{13} {\rm M_{\odot}}$ & $10^{12} {\rm M_{\odot}}$ &  / Cool core \\ 
  \hline

A0085      & 00:41:50.306 & -09:18:11.11 & 0.0556 &$  6.37  \pm  1.00  $&$  5.68  \pm  0.37  $&$  8.13  \pm  0.38  $&$  2.71  \pm  1.83   $ &  Y/S\\
A0400      & 02:57:41.349 & +06:01:36.93 & 0.0240 &$  1.83  \pm  0.39  $&$  1.07  \pm  0.07  $&$  1.36  \pm  0.05  $&$  1.58  \pm  1.07   $ &  N/N\\
IIIZw54    & 03:41:18.729 & +15:24:13.91 & 0.0311 &$  1.91  \pm  0.58  $&$  1.18  \pm  0.08  $&$  1.45  \pm  0.26  $&$  2.79  \pm  1.56   $ &  Y/W\\
A1367      & 11:44:44.501 & +19:43:55.82 & 0.0216 &$  1.76  \pm  0.27  $&$  2.11  \pm  0.14  $&$  2.07  \pm  0.07  $&$  1.72  \pm  0.30   $ &  N/N\\
MKW4       & 12:04:27.660 & +01:53:41.50 & 0.0200 &$  0.50  \pm  0.14  $&$  0.58  \pm  0.04  $&$  0.47  \pm  0.02  $&$  0.63  \pm  0.18   $ &  Y/S\\
ZwCl1215   & 12:17:40.637 & +03:39:29.66 & 0.0750 &$  4.93  \pm  0.98  $&$  4.34  \pm  0.28  $&$  6.10  \pm  0.29  $&$  3.06  \pm  2.30   $ &  Y/N\\
A1650      & 12:58:41.885 & -01:45:32.91 & 0.0845 &$  3.44  \pm  0.66  $&$  4.28  \pm  0.27  $&$  5.09  \pm  0.73  $&$  3.09  \pm  1.43   $ &   Y/W\\
Coma       & 12:59:45.341 & +27:57:05.63 & 0.0232 &$  6.55  \pm  0.79  $&$  6.21  \pm  0.40  $&$  8.42  \pm  0.63  $&$  2.23  \pm  1.12   $ &   N/N\\
A1795      & 13:48:52.790 & +26:35:34.36 & 0.0616 &$  3.41  \pm  0.63  $&$  4.46  \pm  0.29  $&$  5.11  \pm  0.14  $&$  2.44  \pm  0.81   $ &   Y/S\\
MKW8       & 14:40:42.150 & +03:28:17.87 & 0.0270 &$  0.62  \pm  0.12  $&$  1.10  \pm  0.07  $&$  0.80  \pm  0.12  $&$  0.52  \pm  0.39   $ &   N/N\\
A2029      & 15:10:55.990 & +05:44:33.64 & 0.0767 &$ 14.70  \pm  2.61  $&$  6.82  \pm  0.44  $&$ 13.35  \pm  0.53  $&$  4.49  \pm  0.80   $ &   Y/S\\
A2052      & 15:16:44.411 & +07:01:12.57 & 0.0348 &$  1.39  \pm  0.28  $&$  2.03  \pm  0.13  $&$  1.86  \pm  0.10  $&$  2.84  \pm  0.33   $ &   Y/S\\
MKW3S      & 15:21:50.277 & +07:42:11.77 & 0.0450 &$  1.45  \pm  0.34  $&$  2.29  \pm  0.15  $&$  2.13  \pm  0.09  $&$  1.30  \pm  1.76   $ &   Y/S\\
A2065      & 15:22:29.082 & +27:43:14.39 & 0.0721 &$ 11.18  \pm  1.78  $&$  3.35  \pm  0.22  $&$  7.66  \pm  1.44  $&$  3.04  \pm  1.53   $ &   N/W\\
A2142      & 15:58:19.776 & +27:14:00.96 & 0.0899 &$  7.36  \pm  1.25  $&$ 10.26  \pm  0.66  $&$ 13.76  \pm  0.73  $&$  4.04  \pm  2.22   $ &   Y/W\\
A2147      & 16:02:16.305 & +15:58:18.46 & 0.0351 &$  4.44  \pm  0.67  $&$  3.63  \pm  0.23  $&$  5.04  \pm  0.53  $&$  2.37  \pm  1.76   $ &   N/N\\
A2199      & 16:28:37.126 & +39:32:53.29 & 0.0302 &$  2.69  \pm  0.42  $&$  2.64  \pm  0.17  $&$  2.97  \pm  0.30  $&$  2.26  \pm  0.06   $ &   Y/S\\
A2255      & 17:12:54.538 & +64:03:51.46 & 0.0800 &$  7.13  \pm  1.38  $&$  4.08  \pm  0.26  $&$  7.11  \pm  0.33  $&$  2.99  \pm  1.79   $ &   N/N\\
A2589      & 23:23:56.772 & +16:46:33.19 & 0.0416 &$  3.03  \pm  0.75  $&$  1.88  \pm  0.12  $&$  2.54  \pm  0.17  $&$  1.77  \pm  2.04   $ &   Y/W\\

\hline
\end{tabular} 
\label{t:sample}
\hspace*{0.3cm}{\footnotesize Note: The cluster mass, $M_{\rm
    500,M-M_{gas}}$, is derived from the $M_{500}-M_{\rm gas,500}$
  relation as tabulated in Lagan\'a et al. (2011), and only used for
  comparison with the cluster mass, $M_{500}$, derived from the
  ``harmonic'' velocity dispersion. ``S'', ``W'', and ``N'' denote
  strong cool-core, weak cool-core, and noncool-core clusters (see
  Hudson et al. 2010). }
\end{table*}
\end{center}

\begin{center}
\begin{table*}
  \caption{Slopes and characteristic magnitudes of the double Schechter
    function fit to the GLF, gas-mass fraction, and stellar mass
    fraction.}
\begin{tabular}{ccccccc}
  \hline 
Name & $\alpha_{1}$ & $M_{*,1}$ & $\alpha_{2}$ & $M_{*,2}$
  & $f_{\rm gas,500}$ & $f_{\rm *,500}$ \\
\hline
A0085      &$ -1.484 \pm  0.045 $&$ -21.87 \pm   0.71 $&$ -1.945 \pm  0.072 $&$ -20.30 \pm   0.05 $&$  0.128  \pm  0.010  $&$ 0.0115  \pm 0.0012  $\\
A0400      &$ -1.194 \pm  0.057 $&$ -22.24 \pm   0.11 $&$ -1.821 \pm  0.027 $&$ -18.94 \pm   0.06 $&$  0.075  \pm  0.008  $&$ 0.0240  \pm 0.0039  $\\
IIIZw54    &$ -1.289 \pm  0.120 $&$ -22.00 \pm   0.70 $&$ -1.651 \pm  0.028 $&$ -18.42 \pm   0.01 $&$  0.076  \pm  0.013  $&$ 0.0239  \pm 0.0039  $\\
A1367      &$ -1.184 \pm  0.029 $&$ -22.72 \pm   0.81 $&$ -1.922 \pm  0.062 $&$ -18.98 \pm   0.04 $&$  0.118  \pm  0.009  $&$ 0.0247  \pm 0.0028  $\\
MKW4       &$ -1.666 \pm  0.073 $&$ -20.62 \pm   0.35 $&$ -1.802 \pm  0.024 $&$ -19.65 \pm   0.10 $&$  0.094  \pm  0.013  $&$ 0.0233  \pm 0.0040  $\\
ZwCl1215   &$ -1.698 \pm  0.013 $&$ -22.02 \pm   0.11 $&     --- & --- &$  0.124  \pm  0.013  $&$ 0.0143  \pm 0.0017  $\\
A1650      &$ -1.271 \pm  0.051 $&$ -23.10 \pm   0.60 $&     --- & --- &$  0.148  \pm  0.018  $&$ 0.0217  \pm 0.0026  $\\
Coma       &$ -1.303 \pm  0.025 $&$ -22.34 \pm   0.36 $&$ -1.935 \pm  0.019 $&$ -19.76 \pm   0.08 $&$  0.129  \pm  0.009  $&$ 0.0201  \pm 0.0018  $\\
A1795      &$ -1.327 \pm  0.006 $&$ -21.48 \pm   0.61 $&$ -1.847 \pm  0.032 $&$ -20.17 \pm   0.02 $&$  0.150  \pm  0.014  $&$ 0.0182  \pm 0.0022  $\\
MKW8       &$ -1.145 \pm  0.074 $&$ -20.55 \pm   0.45 $&$ -1.959 \pm  0.028 $&$ -19.32 \pm   0.02 $&$  0.128  \pm  0.016  $&$ 0.0259  \pm 0.0031  $\\
A2029      &$ -1.227 \pm  0.010 $&$ -21.83 \pm   0.86 $&     --- & --- &$  0.091  \pm  0.008  $&$ 0.0065  \pm 0.0007  $\\
A2052      &$ -1.313 \pm  0.005 $&$ -21.74 \pm   0.81 $&$ -1.964 \pm  0.014 $&$ -19.68 \pm   0.01 $&$  0.134  \pm  0.014  $&$ 0.0254  \pm 0.0030  $\\
MKW3S      &$ -1.422 \pm  0.033 $&$ -22.62 \pm   0.49 $&$ -1.822 \pm  0.036 $&$ -19.81 \pm   0.03 $&$  0.147  \pm  0.017  $&$ 0.0268  \pm 0.0035  $\\
A2065      &$ -1.061 \pm  0.029 $&$ -22.01 \pm   0.26 $&$ -1.643 \pm  0.015 $&$ -20.46 \pm   0.01 $&$  0.069  \pm  0.008  $&$ 0.0065  \pm 0.0006  $\\
A2142      &$ -1.271 \pm  0.014 $&$ -21.87 \pm   0.10 $&$ -1.495 \pm  0.094 $&$ -20.38 \pm   0.01 $&$  0.187  \pm  0.017  $&$ 0.0114  \pm 0.0011  $\\
A2147      &$ -1.304 \pm  0.009 $&$ -21.52 \pm   0.12 $&$ -1.879 \pm  0.015 $&$ -19.91 \pm   0.01 $&$  0.114  \pm  0.010  $&$ 0.0154  \pm 0.0015  $\\
A2199      &$ -1.154 \pm  0.031 $&$ -21.44 \pm   0.34 $&$ -2.024 \pm  0.014 $&$ -19.65 \pm   0.01 $&$  0.110  \pm  0.010  $&$ 0.0177  \pm 0.0017  $\\
A2255      &$ -1.419 \pm  0.024 $&$ -23.02 \pm   0.98 $&     --- & --- &$  0.100  \pm  0.010  $&$ 0.0095  \pm 0.0011  $\\
A2589      &$ -1.085 \pm  0.013 $&$ -21.71 \pm   0.17 $&$ -1.838 \pm  0.021 $&$ -19.89 \pm   0.01 $&$  0.084  \pm  0.011  $&$ 0.0169  \pm 0.0023  $\\
\hline
\end{tabular} 
\label{t:glf}
\hspace*{0.3cm}{\footnotesize  }
\end{table*}
\end{center}

\begin{table*}
\begin{center}
  \caption{Comparison of the best fits of the gas-mass fraction,
    stellar mass fraction, and baryon-mass fraction as a function of
    the cluster total mass between different samples.}
\begin{tabular}{lll}
  \hline 
  Sample & Best fit & Intrinsic scatter \\
  \hline
  19clusters & $f_{\rm
gas,500}=10^{-(0.93\pm0.32)}( {M_{500}}/[{10^{14}{\rm M_{\odot}}}]
)^{-(0.03\pm 0.64)}$ & $(26\pm \;\,4)$\%\\
  19clusters-A2029-A2065 & 
$f_{\rm gas,500}=10^{-(1.10\pm0.16)}(
{M_{500}}/[{10^{14}{\rm M_{\odot}}}] )^{0.38\pm 0.36}$ & $(23\pm
17)$\%\\ 
  19clusters and Sun+09& $f_{\rm
gas,500}=10^{-(1.07\pm0.02)}( {M_{500}}/[{10^{14}{\rm M_{\odot}}}]
)^{0.30\pm 0.07}$ & $(26\pm\;\, 8)$\% \\
Pratt+09               & $f_{\rm 
gas,500}=10^{-(1.029\pm0.013)h^{-1.5}(z)}( {M_{500}}/[{2\times 10^{14}{\rm M_{\odot}}}] )^{(0.21\pm 0.03)h^{-1.5}(z)}$ & $(12\pm\;\, 2)$\% \\
  Sun+09 (Tier 1 groups + clusters)& $f_{\rm
gas,500}=(0.0724\pm0.0078)h^{-1.5}_{73}( {M_{500}}/[{10^{13}h^{-1}_{73}{\rm M_{\odot}}}]
)^{0.093\pm 0.031}$ &  --- \\
  Andreon 2010 & $f_{\rm
gas,500}=10^{-(0.97\pm0.02)}( {M_{500}}/[{10^{14.5}{\rm M_{\odot}}}]
)^{0.15\pm 0.03}$ & $(14\pm \;\,2)$\%\\
\hline
  19clusters & $f_{\rm
*,500}=10^{-(1.53\pm0.05)}( {M_{500}}/[{10^{14}{\rm M_{\odot}}}]
)^{-(0.49\pm 0.09)}$ & $(20\pm \;\,4)$\%\\
  19clusters-A2029-A2065 & 
$f_{\rm *,500}=10^{-(1.56\pm0.05)}(
{M_{500}}/[{10^{14}{\rm M_{\odot}}}] )^{-(0.39\pm 0.09)}$ & $(15\pm
29)$\%\\ 
 Lin+03     & $f_{\rm *,500}=0.0164^{+0.0010}_{-0.0090}\left(
{M_{500}}/[{3\times10^{14}{\rm M_{\odot}}}] \right)^{-(0.26\pm 0.09)}$ & ---\\ 
 Gonzalez+07  & $f_{\rm
*,500}=10^{7.57\pm0.08)}{M_{500}}^{-(0.64\pm 0.13)}$ & --- \\
 Giodini+09 & $f_{\rm *,500}=(0.050\pm0.001)(
{M_{500}}/[{5\times10^{13}{\rm M_{\odot}}}] )^{-(0.26\pm 0.09)}$ & 0.35~dex\\ 
 Giodini+09 and Lin+03 & $f_{\rm *,500}=(0.050\pm0.001)(
{M_{500}}/[{5\times10^{13}{\rm M_{\odot}}}] )^{-(0.37\pm 0.04)}$ & 0.50~dex\\ 
  Andreon 2010 & $f_{\rm
*,500} \propto( {M_{500}}/[{10^{14.5}{\rm M_{\odot}}}]
)^{-(0.55\pm 0.08)}$ & $(41\pm \;\,5)$\%\\
 \hline
  19clusters & $f_{\rm
b,500}=10^{-(0.73\pm0.24)}( {M_{500}}/[{10^{14}{\rm M_{\odot}}}]
)^{-(0.30\pm 0.48)}$ & $(45\pm \;\,6)$\%\\
  19clusters-A2029-A2065 & 
$f_{\rm b,500}=10^{-(0.96\pm 0.25)}(
{M_{500}}/[{10^{14}{\rm M_{\odot}}}] )^{0.22\pm 0.57}$ & $(29\pm
14)$\%\\ 
 Lin+03     & $f_{\rm b,500}=0.148^{+0.005}_{-0.004}\left(
{M_{500}}/[{3\times10^{14}{\rm M_{\odot}}}] \right)^{0.148\pm 0.040}$ & ---\\ 
 Giodini+09 & $f_{\rm b,500}=(0.123\pm0.003)(
{M_{500}}/[{2\times10^{14}{\rm M_{\odot}}}] )^{0.09\pm 0.03}$ & ---\\ 
\hline
\end{tabular} 
\label{t:fit}
\end{center}
\hspace*{0.3cm}{\footnotesize }
\end{table*}

\clearpage 

\clearpage

\end{document}